# A Register Allocation Algorithm in the Presence of Scalar Replacement for Fine-Grain Configurable Architectures [*]

Nastaran Baradaran and Pedro C. Diniz
University of Southern California / Information Sciences Institute
Marina del Rey, California 90292, U.S.A.
{nastaran, pedro}@isi.edu

## Abstract

*The aggressive application of scalar replacement to array references substantially reduces the number of memory operations at the expense of a possibly very large number of registers. In this paper we describe a register allocation algorithm that assigns registers to scalar replaced array references along the critical paths of a computation, in many cases exploiting the opportunity for concurrent memory accesses. Experimental results, for a set of image/signal processing code kernels, reveal that the proposed algorithm leads to a substantial reduction of the number of execution cycles for the corresponding hardware implementation on a contemporary Field-Programmable-Gate-Array (FPGA) when compared to other greedy allocation algorithms, in some cases, using even fewer number of registers.*

## 1. Introduction

*Scalar replacement* or *register promotion* is an effective technique for eliminating external memory accesses for the data that is repeatedly accessed throughout a computation. This technique, geared towards array variables, enables a compiler to replace the repeatedly accessed array references by scalar references. Mapping these scalars to hardware registers eliminates the memory operations associated with fetching/storing of the values, while making them readily available for future use. This transformation is particularly suited for loop-based memory-intensive computations, such as those arising in common image and signal processing code kernels, where there are substantial opportunities for both *input* and *output* data reuse.

The aggressive application of *scalar replacement* however may require a large number of registers, limiting the application of this technique. As such, fine-grain configurable architectures, such as Field-Programmable-Gate-Arrays (FPGAs), offer an ideal context for applying the scalar replacement to image and signal processing applications. These architectures have a large yet *limited* number of available registers which can be organized freely, as well as storage structures organized as RAM blocks with programmable bit-widths and flexible number of access ports. A compiler can exploit scalar replaced array references by explicitly mapping and managing the corresponding scalars to a combination of registers and RAM blocks [2].

In this paper we describe several algorithms for the allocation of registers to scalar variables resulting from the application of scalar replacement to array references in perfectly nested loops. We describe and evaluate two *greedy* allocation algorithms based on cost/benefit metrics and propose a novel *critical-path-aware* allocation algorithm. The proposed algorithm allocates registers to references along *cuts* of the critical path of the computation, ensuring that the eliminated memory accesses lead to a reduction of the computation's execution cycles and wall-clock execution time.

We evaluate the performance for the various algorithms using a small set of image/signal processing code kernels. The results reveal that the proposed algorithm is effective in allocating registers to the scalar replaced array references in the code, therefore reducing the number of execution cycles of each computation. In some cases the *critical-path-aware* algorithm reduces the overall execution cycles, as well as the overall execution time, using the same or even fewer number of registers than other greedy algorithms.

In the rest of this paper, section 2 describes background and related work. Sections 3 and 4 formalize our register allocation problem along with the description of the proposed *critical-path-aware* algorithm. We present experimental results in section 5 and conclude in section 6.

[*] This work is supported by the National Science Foundation (NSF) under Grant No. 0209228. Any opinions, findings, and conclusions or recommendations expressed in this material are those of the author(s) and do not necessarily reflect the views of the NSF.

## 2. Background and Related Work

We now briefly describe the relevant features of our target configurable architecture as well as the compiler analysis concepts that support the application of scalar replacement. We also survey related work in the context of mapping array variables to these architectures and contrast these efforts with traditional register allocation approaches.

**Configurable Architectures:** Our work targets configurable architectures with storage resources that can be configured in an application-specific fashion. In addition the target architecture also has a large number of resources which can be organized as either computing elements or discrete data registers. As an example, the Xilinx Virtex-II [13] family of FPGAs have a limited set of RAM blocks that can be configured as single- or dual-ported RAM memories given a fixed bit capacity. The PipeRench [6] opts for a computation execution model based on pipelining the data through a fixed set of stripes, each with finite computational and storage elements. The XPP Array [14] uses coarser grained elements connected via a programmable network and exports a low-level execution model that resembles a data-flow. For a given configuration of each node the execution proceeds when the data inputs are available.

A significant difference between these architectures and traditional processors is the absence of a unified address space and underlying hardware mechanisms to enforce data consistency across the various storage structures. Designers must explicitly map high-level program variables to both RAMs and registers and explicitly manage the flow of data between them to enforce data consistency.

**Data Reuse & Scalar Replacement:** Data reuse analysis for array variables in a loop nest relies on the concept of dependence distance. The compiler observes the array reference index functions, in this context affine functions of the enclosing loop index variables, and understands at which loop iterations the same data element is reused.

```
for (i = 0; i < b_i = 100; i++)
 for (j = 0; j < b_j = 20; j++)
  for (k = 0; k < b_k = 30; k++){
   d[i][k] = a[k] * b[k][j];
   e[i][j][k] = c[j] * d[i][k];
  }
```

**Figure 1. Example Code**

In the code example in figure 1 the reference $b[k][j]$ exhibits reuse at the $i$ loop level, as for every iteration of the $i$ loop the same location is accessed for the same values of the $j$ and $k$ iterations. *Scalar replacement* converts array references into scalar variables and then maps them to registers. For $b[k][j]$, one can save the $b_k \times b_j$ accesses to $b[0][0], \ldots, b[b_k - 1][b_j - 1]$ in scalar variables for the first iteration of $i$, and then reuse these values for the subsequent $b_i - 1$ iterations of the $i$ loop. By doing so the implementation eliminates $b_k \times b_j \times (b_i - 1)$ memory accesses at the expense of $b_k \times b_j$ scalar variables.

Researchers have developed several compiler data dependence analysis frameworks for uncovering data reuse for affine references in loop nests [4, 8], and have analytically computed the number of required registers to capture reuse across the various loop levels in a nest [11]. As for code generation, the application of scalar replacement and subsequent mapping to registers can be accomplished by pre-peeling the iterations of the loop where input data needs to be saved in registers, or back-peeling the iterations of the loop where the data needs to be restored to memory. The complete code generation scheme, either using peeling or predication, is beyond the scope of this paper.

**Storage Resource Allocation:** Minimizing the impact of the access to memory has been a long standing problem. Gokhale *et al.* [5] describe an algorithm for the mapping of array variables to external memories in FPGA-based architectures. Weinhardt and Luk [12] describe a limited compiler approach for using RAM blocks to cache the data in contemporary FPGAs. In our own work we have used the same data reuse analysis framework outlined in this paper to explore the area and space trade-offs of using RAM blocks to store scalar replaced variables [2], whereas So and Hall [11] exclusively use registers to cache the data. There has also been extensive work in hierarchical data mapping in order to improve overall performance metrics such as time or power [1, 7, 10].

The classical register allocation problem focuses on the assignment of a finite number of registers to scalar variables only. Given the significance of this problem, and its intractable worst case complexity, many researchers have developed various algorithmic strategies. For example, Briggs *et. al.* [3] describe several graph coloring heuristics whereas Kolson *et. al.* [9] propose a spill minimizing register allocation algorithm for embedded code generation.

Our register allocation approach differs from these efforts in several aspects. First, we use scalar replacement information to select the more profitable array references in order to limit the number of required registers, without limiting the reuse to innermost loop levels. Second, we exploit the data-flow information of the computation to coallocate registers to inputs of the same operation. Finally, and as with other approaches for configurable architectures, our approach and corresponding code generation must explicitly manage the flow of data between registers and RAMs.

## 3. Problem Formulation and Definitions

The register allocation for the scalars generated by an aggressive application of scalar replacement can be formulated as a *Knapsack problem*. In this formulation, an object is an array reference represented by ref$_i$, whereas the size of each object is the number of required registers[1] for a full scalar replacement of a reference and is represented by $\alpha_i$. Furthermore, the value of each object is the potential number of eliminated memory accesses and the size of the register file is the knapsack size. A simple objective function is to eliminate the most memory accesses [4].

This formulation however does not take into account the dependences between references and the opportunities for concurrent data accesses to RAM blocks. If references corresponding to distinct array variables are mapped to different RAMs, accesses to them can proceed concurrently, only incurring the latency of a single access. Considering this concurrency opportunity, we formulate the register allocation problem for scalar replaced array references as finding a register allocation that minimizes the completion time for the computation in a loop nest.

To capture the notion of execution time, we abstract the computation in a loop nest as a collection of data-flow graphs (DFG). In this abstraction ref$_i$ and op$_i$ represent various array references and operations in the DFG, while $Lat(op_i)$ captures the latency of a specific numeric operation or a memory access. We further assume the latencies of the numeric operations to be known and the latency of a memory access for a specific array reference to be either 0 or 1, depending on whether the array element is mapped to a register or to a RAM block. Given a DFG, we define the latency of a path $p_k$ as $Lat(p_k) = \sum_{op_i \in p_k} Lat(op_i)$ and determine the *Critical Path(s)* (CP) of a DFG as the path(s) with the highest latency. Finally, we define the execution time $T_{exec}$ of a DFG as the maximum latency across its paths, i.e., $T_{exec} = Max^{p_k \in Paths} Lat(p_k)$, or simply $T_{exec} = Lat(CP)$. Given these definitions, we wish to determine a register allocation that minimizes the memory access portion of $T_{exec}$ for the entire execution of the loop, subject to the available number of registers $NR$.

In order to reduce the overall execution time, all the critical paths in a DFG should be reduced. Improving only a subset of the CPs would just consume the resources without having any effect on the overall computation time. To address this issue we introduce the *Critical Graph* (CG) as a subgraph of DFG including all of its CPs. We also define a *Cut* of the Critical Graph (CG) as a minimal subset of its reference nodes, such that their removal would disconnect all the paths in the CG.[2] Therefore, in order to improve the $T_{exec}$, all the references in a *Cut* need to be stored in registers.

Figures 2(a) and (b) depict the DFG and CG, along with the set of possible cuts, for the example code in figure 1. In terms of full scalar replacement, the references a,b,c,d and e would require $\alpha_a = 30$, $\alpha_b = 600$, $\alpha_c = 20$, $\alpha_d = 30$, and $\alpha_e = 1$ registers respectively. Given a limit of 64 registers, we can not possibly accommodate all scalar variables for all references. If we assign the scalar variables generated by a[k] to 30 registers while keeping the scalar variables of b[k][j] in a RAM, for each a[k] * b[k][j] operation we would need to read the data for b[k][j] from the RAM. The registers assigned to a[k] would not be used effectively since the execution of the operation would need to stall until the values corresponding to b[k][j] would be retrieved from RAM. Instead we could allocate the available 30 registers to both $a$ and $b$ in order to improve the data access. Even if we could not fully assign all the scalar variables for b[k][j], at least for a subset of the operations in the $k$ loop, both operations would use the data in registers.

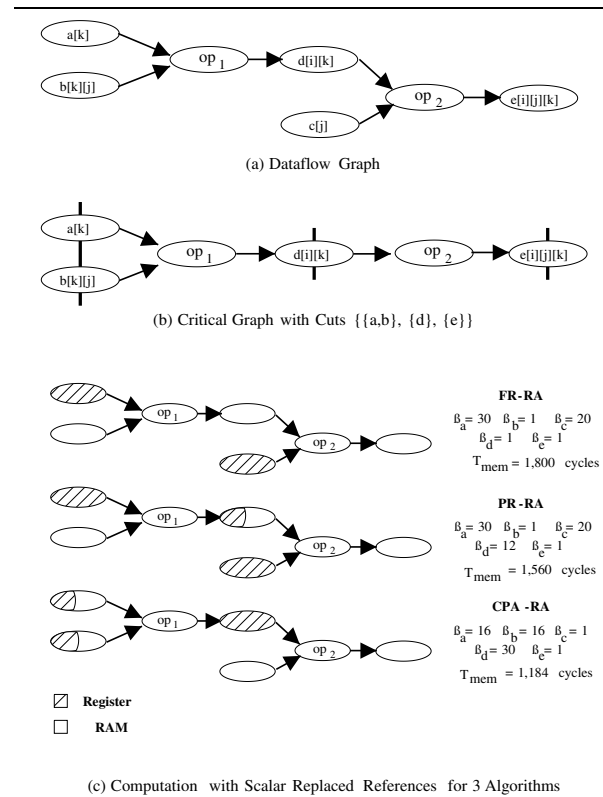

**Figure 2. Stages of Allocation Algorithm.**

---

1  The techniques to calculate this number have been extensively addressed in [11].
2  A simple algorithm to find a cut of a graph consists of iteratively selecting a node of the graph and eliminating all its ancestors and descendants until no more nodes are left in the graph. In the worst-case, finding all the cuts of a graph is exponential.

```
Inputs: α_i, BC(ref_i), NR, NA
Output: β_i: Number of registers assigned to array ref_i

Knapsack Solution Full and Partial Reuse {
 for 1 ≤ i ≤ NA do set β_i = 1
 if (∑_{i=1}^{NA} α_i ≤ NR) then
  for (i=1 to i = NA)
   β_i = α_i
 else
 // Variant 1. Full Reuse Allocation
  sort array references {ref_1,...,ref_NA}
  based on the descending value of BC.
  for each ref ref_i in the sorted list do
   if (NR ≥ α_i) then
    β_i = α_i
    NR = NR - α_i
   end if
  end for
 // Variant 2. Partial Reuse Allocation
  for the first ref_j where β_j = 1 do
   β_j = NR
}
```

**Figure 3. Full Reuse and Partial Reuse Register Allocation Algorithms (FR-RA & PR-RA).**

```
Input: α_i, NR, dfg Data Flow Graph.
Output: β_i: Number of registers assigned to array ref_i

CriticalPathAware{
 for 1 ≤ i ≤ NA do set β_i = 1
 while (NR > 0) do
  cg = Make_CG(dfg);
  cg_cuts = Find_Cuts(cg);
  Find_Req_Reg(cg_cuts);
  best_cut : Element of the cg_cuts with the min Req_Reg
  RR = ∑_{i=1}^{nodes∈best_cut} α_i
  if (RR ≤ NR) then
   for (∀ref_i ∈ best_cut) do
    β_i = α_i
    NR -= RR
    RR = 0
  end if
  if (RR ≠ 0 && 0 < NR) then
   for (∀ref_i ∈ best_cut) do
    β_i = NR/(Num of references in best_cut)
  end if
}
```

**Figure 4. Critical-Path-Aware Register Allocation Algorithm (CPA-RA).**

## 4. Allocation Algorithms

We now present several *greedy* algorithms to tackle the *knapsack* problem as formulated in section 3. In the description of the algorithm we denote $BC(\text{ref}_i)$ as the benefit/cost metric defined as the ratio of saved memory accesses over the number of required registers for reference $\text{ref}_i$. We denote the maximum number of available registers and number of array references by $NR$ and $NA$ respectively. Finally, $\beta_i$ indicates the number of registers that the algorithm assigns to reference $\text{ref}_i$.

The first variant, named *Full Reuse Register Allocation (FR-RA)*, starts by assigning one register to each array reference to render the computation feasible. It then uses the value of $BC(\text{ref}_i) = Save(\text{ref}_i)/\alpha_i$ to greedily assign the available registers to the data references that yield the best benefit/cost ratio. For each reference $\text{ref}_i$, if possible, the algorithm assigns $\alpha_i$ registers corresponding to fully exploiting the data reuse for that reference. This proceeds until the algorithm exhausts all the available registers leading to an assignment of either $\alpha_i$ or 1 to $\beta_i$.

This simple greedy algorithm might leave some registers unallocated, as upon termination the remaining number of registers might not be enough to satisfy a value of $\alpha_i$. In the variant 2 of the algorithm, named *Partial Reuse Register Allocation (PR-RA)*, we allow the assignment of the extra registers to the next reference in the sorted list. For this reference the implementation exploits partial data reuse, as it assigns $\beta_{ref}$ registers with $1 \leq \beta_{ref} < \alpha_{ref}$.

Clearly, these algorithms do not attempt to reduce the computation time. To address this issue, we use a *greedy* approach named *Critical Path Aware Register Allocation (CPA-RA)* algorithm. This last variant calculates the value of $\beta_i$ for each $\text{ref}_i$ by finding the most valuable references on the Critical Path (CP) and distributing the available registers among them, with the objective of minimizing the memory access time along the CP. In order to determine the critical path of a set of DFGs, the algorithm assumes a specific scheduling implied by the mapping of array variables to RAM blocks and the limited computing resources.

CPA-RA starts by constructing the Data-Flow Graph (DFG) for the computations in the loop body. It then extracts the Critical Graph and finds all the possible cuts of the CG. After calculating the number of registers required to fully accommodate each cut ($\Omega = \sum_{i=1}^{nodes \in cut} \alpha_i$), the algorithm selects the cut with the min $\Omega$. For the selected cut, if possible, the algorithm assigns $\Omega$ registers corresponding to fully exploiting the data reuse for the references of the cut. Otherwise it divides (equally) the available registers between the references. The algorithm repeats this process until it consumes all the available registers. The complexity of the algorithm is a function of the number of critical paths and their memory accesses, and therefore is exponential in the worst-case. However in practice, and in our experiments, the CG is generally so small that this fact is not a concern.

We now illustrate the application of these three algorithms to the example code in figure 1 using 64 available registers. For this code, the benefit/cost of the references yield the values $BC(\text{a}) = 1999$, $BC(\text{b}) = 99$, $BC(\text{c}) = 2999$, $BC(\text{d}) = 1900$, and $BC(\text{e}) = 1$. The FR-RA algorithm assigns the available 64 registers to the references in the order of c, a, d, b, e resulting in an assignment of $\beta = \{20, 30, 1, 1, 1\}$. The PR-RA algorithm assigns registers in the same order but since there are 11 registers left, it assigns them to the d array reference resulting in the as-signment $\beta = \{20, 30, 12, 1, 1\}$. Finally, the CPA-RA algorithm first selects cut $\{d\}$ due to its minimum number of required registers and assigns 30 registers to this reference, thereby reducing the length of the CG by one node (memory access). In a second iteration, the algorithm picks the cut $\{a, b\}$ and assigns the remainder of the registers equally to references a and b.

Figure 2(c) illustrates the register distribution for the various arrays of figure 1 as a result of applying the above algorithms. Considering the loop bounds and the latency of each RAM access, under a serial execution, the code resulting from the application of the FR-RA algorithm would exhibit 1800 cycles for memory operations whereas using the PR-RA algorithm it would exhibit only 1560 cycles, since 12 out of the 30 iterations of $k$ have only 2 memory accesses. For the CPA-RA algorithm, iterations have either 2 or 1 memory accesses, due to the full scalar replacement of d and a partial scalar replacement of a and b. As a result a total of 1184 cycles are devoted to memory operations. It is important to notice that, for this example, CPA-RA substantially reduces the cycles devoted to memory operations using the exact same register resources.

## 5. Experimental Results

We validated the register allocation algorithm for a set of six signal and image processing code kernels. The Finite-Impulse-Response (FIR) and Decimation FIR filter (Dec-FIR) code kernels compute a convolution of a 1024-long vector of 16-bit values against a 52 and 128-long sequence of coefficients, with and without a decimation factor of 2 respectively. The MAT kernel performs a $16 \times 16$ matrix-matrix multiplication. The IMI kernel computes the interpolation of two grey-scaled $8 \times 6$ images for 30 intermediate image values. The PAT kernel finds the various occurrences of an 80-character long string pattern in a 1024 length string. Finally, BIC computes a Binary-Image-Correlation between a $8 \times 8$ template image and successively overlapping regions of a larger $64 \times 64$ image. With the exception of MAT and BIC, which are structured as a 3- a 4-deep nested loops respectively, all kernels are structured as 2-deep loop nests with compile-time known bounds.

For each kernel, written in C, we applied *scalar replacement* at the source C level and then converted the transformed C codes to behavioral VHDL. To decouple the experiment from the code generation complexity issues of *scalar replacement* due to the use of *loop peeling*, we opted to use the same structure of control (in terms of loops and peeled sections) for all of the code versions. Next we converted the behavioral descriptions of the codes into a structural VHDL design using Mentor Graphics' Monet™ high-level synthesis tool. We then used Synplify Pro 6.2 and Xilinx ISE 4.1i tool sets for logic synthesis and Place-and-Route (P&R) targeting a Xilinx Virtex™ XCV 1K BG560 device. After P&R we extracted the real area and clock rate for each design and used the number of cycles derived from the simulation to calculate wall-clock execution time.

In these experiments we imposed a maximum limit of 64 registers each implementation uses to capture data reuse. In practice this limit must be imposed by the compiler as part of a global resource allocation policy, orthogonal to these experiments. For each code kernel we derived three designs, respectively v1, v2 and v3, reflecting the three register allocation algorithm variants FR-RA, PR-RA and CPA-RA described in section 4.

Table 1 depicts the results for the register allocation and corresponding hardware designs. The third and forth columns indicate the number of registers required by each array reference for a full scalar replacement, and the registers allocated by the algorithms, respectively. The fifth column presents the number of execution cycles, indicating the percentage reduction with respect to the base code version v1. The sixth column presents the attained clock period for the hardware design in nano-seconds, as extracted after P&R. The seventh column presents the wall-clock time for the execution of the computation which takes into account the attained clock rate. The execution time data is used to calculate the speedup of the implementations with respect to the base version. Finally the last two columns present the resources used by each design in terms of slices (out of a maximum of $12, 288$) and number of RAM blocks.

In terms of the register allocation algorithms, code versions v2 use substantially more registers than the corresponding versions v1, in an attempt to exploit partial data reuse. Versions v3 use almost all the available registers as they evenly distribute the number of registers among the operations on the critical path.

As expected, using more registers leads to a reduction of the number of RAM accesses and hence to a reduction in the number of execution cycles. The figures in column 5 show consistently positive gains with an average percentage improvement of 7.7% and 21.3% for versions v2 and v3 respectively. In some cases, such as Dec-FIR and PAT, using more registers in v2 does not lead to a reduction in the number of cycles as the inputs to the same operations are located in distinct types of storage. In fact because the control for these designs is more complex than the base version v1 there is an increase in the clock period leading to an overall performance degradation as revealed by column 7.

The CPA-RA algorithm mitigates this problem by allocating registers to references that always decrease the number of clock cycles. The results reveal that, even though there is a noticeable clock degradation for the more complex v3 designs, the reduction in the number of clock cycles compensates for this clock rate degradation, improving the results of v2 for all cases but MAT and BIC. For configurable architectures where the clock rate is fixed regardless of the design complexity, the results would yield performance improvements for all code variants as derived from the reduction of the number of clock cycles.

| Kernel | Code Version | Required S.R. Registers | Number of Registers Distribution | Number of Registers Total | Total Execution Cycle Count | | Clock Period | Execution Time($\mu$s) | | FPGA Slices Count | FPGA Slices Occupancy | RAMs Used |
|---|---|---|---|---|---|---|---|---|---|---|---|---|
| FIR | v1 | 1, 52, 51 | 1, 52, 1 | (54) | 160, 769 | (—) | 34.65 | 5, 571 | (—) | 1, 100 | (8%) | 1 |
|  | v2 |  | 1, 52, 11 | (64) | 141, 334 | (12.0%) | 37.68 | 5, 325 | (−4.4%) | 1, 405 | (11%) | 1 |
|  | v3 |  | 1, 12, 51 | (64) | 97, 291 | (39.5%) | 41.06 | 3, 994 | (28.3%) | 1, 286 | (10%) | 2 |
| Dec-FIR | v1 | 1, 128, 127 | 1, 1, 1 | (3) | 777, 490 | (—) | 37.48 | 29, 140 | (—) | 672 | (5%) | 2 |
|  | v2 |  | 1, 62, 1 | (64) | 777, 490 | (0.0%) | 38.34 | 29, 809 | (−2.3%) | 1, 696 | (13%) | 2 |
|  | v3 |  | 1, 1, 62 | (64) | 716, 139 | (7.9%) | 41.33 | 29, 598 | (−1.5%) | 2, 003 | (16%) | 2 |
| IMI | v1 | 1, 48, 48 | 1, 48, 1 | (50) | 4, 788 | (—) | 27.68 | 132.9 | (—) | 301 | (2%) | 1 |
|  | v2 |  | 1, 48, 15 | (64) | 3, 800 | (20.6%) | 28.39 | 107.8 | (18.8%) | 394 | (3%) | 1 |
|  | v3 |  | 1, 31, 31 | (63) | 2, 849 | (40.7%) | 29.50 | 83.7 | (36.8%) | 691 | (5%) | 2 |
| MAT | v1 | 1, 16, 256 | 1, 16, 1 | (18) | 14, 625 | (—) | 28.92 | 423.0 | (—) | 311 | (2%) | 1 |
|  | v2 |  | 1, 16, 47 | (64) | 13, 809 | (5.5%) | 30.10 | 416.0 | (1.73%) | 624 | (5%) | 1 |
|  | v3 |  | 1, 16, 47 | (64) | 13, 809 | (5.5%) | 30.10 | 416.0 | (1.73%) | 624 | (5%) | 1 |
| PAT | v1 | 1, 80, 79 | 1, 1, 1 | (3) | 250, 879 | (—) | 25.26 | 6, 337 | (—) | 184 | (1%) | 2 |
|  | v2 |  | 1, 62, 1 | (64) | 250, 879 | (0%) | 29.82 | 7, 481 | (−18%) | 1, 086 | (8%) | 2 |
|  | v3 |  | 1, 1, 62 | (64) | 186, 368 | (25.7%) | 29.21 | 5, 444 | (14.1%) | 433 | (3%) | 2 |
| BIC | v1 | 1, 64, 512 | 1, 56, 1 | (58) | 633, 024 | (—) | 69.69 | 44, 115 | (—) | 7, 524 | (61%) | 1 |
|  | v2 |  | 1, 56, 7 | (64) | 579, 720 | (8.4%) | 67.51 | 39, 137 | (11.3%) | 7, 307 | (59%) | 1 |
|  | v3 |  | 1, 56, 7 | (64) | 579, 720 | (8.4%) | 67.51 | 39, 137 | (11.3%) | 7, 307 | (59%) | 1 |

**Table 1. Analysis and experimental results.**

Overall, code versions v2 exhibit an unimpressive average wall-clock time gain of 2.65% whereas the code versions v3 yield a respectable 15.1% gain even for an average clock rate loss of 8.7%. It is worth to notice that CPA-RA improves the performance of versions v3 over v2, for an average 15% and 12.5% for clock cycles and wall-clock time respectively. This improvement is achieved in many cases with little or no additional registers and no significant increase in the used number of slices, making the proposed CPA-RA a very effective register allocation algorithm for this class of configurable computing architectures.

## 6. Conclusion

Emerging configurable architectures exhibit a rich set of storage and computing resources which must be explicitly managed by compilers for maximum efficiency. In this paper we have described a register allocation algorithm for scalar variables resulting from the aggressive application of scalar replacement. We proposed a critical-path-aware allocation strategy that exploits the internal registers and RAM blocks parallel accesses. We showed that this algorithm leads to substantial performance gains over common register allocation strategies.